# QBism Is Not So Simply Dismissed

Ali Barzegar[1]

Ludwig Maximilian University, Munich

## Abstract

QBism is one of the main candidates for an epistemic interpretation of quantum mechanics. According to QBism, the quantum state or the wavefunction represents the subjective degrees of belief of the agent assigning the state. But, although the quantum state is not part of the furniture of the world, quantum mechanics grasps the real via the Born rule which is a consistency condition for the probability assignments of the agent. In this paper, we evaluate the plausibility of recent criticism of QBism. We focus on the consequences of the subjective character of the quantum state, the issue of realism and the problem of the evolution of the quantum state in QBism. In particular, drawing upon Born's notion of invariance as the mark of the real, it is argued that there is no essential difference between Einstein's program of 'the real' and QBists' realism. Also, it will be argued that QBism can account for the unitary evolution of the quantum state.

## 1. Introduction

In recent times we have been witnessing a turn towards epistemic interpretations of quantum mechanics. According to all these interpretations the quantum state somehow represents the knowledge, information or beliefs of the user of quantum mechanics. QBism is one such interpretation. Other notable epistemic interpretations include Rovelli (1996); Zeilinger (1999, 2005); Healey (2012); Friederich (2015); Mueller (2017); Brukner (2017) and Boge (2018). In this respect

[1] . barzegar11@gmail.com

it should be noted that there are in general three options for interpreting quantum mechanics (Leifer 2011):

1. The quantum state is epistemic and there is some underlying ontic state.
2. The quantum state is epistemic but there is no deeper underlying reality.
3. The quantum state is ontic.

Consequently, it is worth examining the prospects for the epistemic interpretation as a viable interpretation of quantum mechanics. In this article, we focus on the status of QBism as a candidate among them.

QBism is an interpretation of quantum mechanics and more broadly a philosophy of science inspired by the implications of quantum information theory (Fuchs 2002). Its main proponents are Christopher Fuchs, Rüdiger Schack and David Mermin (see e.g. Mermin (2012, 2014); Fuchs (2010); Fuchs, Mermin, and Schack (2014)).

According to QBism, the quantum state or the wavefunction represents the subjective degrees of belief of the agent assigning it. The quantum state is not part of the furniture of the world. On the other hand, the quantum system is an objective element of reality and part of the stuff of the world. Quantum mechanics is about the actions or measurements of the agent on the quantum system and the reactions she receives from the quantum system in the form of measurement outcomes. For QBists, as Fuchs et al. (2014) put it, 'quantum mechanics is a tool anyone can use to evaluate, on the basis of one's past experience, one's probabilistic expectations for one's subsequent experience'. In other words, quantum mechanics provides the agent with coherence conditions for assigning probabilities (Fuchs and Schack 2018). This coherence condition is fulfilled through the Born rule. Quantum mechanics advises the agent to assign and update her probabilities so that the Born rule is satisfied. This normative character of the Born rule is an important tenet of QBism. Being normative, the Born rule neither describes how nature actually is, nor does it tell nature what to do. It is rather an ideal relation an agent strives for in her probability assignments.

The importance of measurement for QBism links it with Bohr and Copenhagen-type interpretations.[2] This centrality of measurement is an essential element of what is known as operationalism. However, for QBists, quantum theory is not just about measurements or a tool to predict measurement outcomes. QBism is a realist rather than an instrumentalist program as will be explained in the course of the paper.

Its proponents believe that QBism solves the longstanding puzzles of quantum mechanics such as the collapse of the wave function and nonlocality. However, QBism is not without its critics. There have been critical stands toward QBism in the literature (see e.g. Timpson (2008); Bacciagaluppi (2014)). Recently, Brown (2017) has put forward a critical review of QBism which improves upon some points in Timpson's 2008 critique and also offers new lines of attack. The aim of this paper is to evaluate the plausibility of Brown's arguments against QBism.

The structure of the paper is as follows. In section 2, Brown's argument for considering the quantum state as part of the stuff of the world is evaluated and rejected. Section 3 offers an answer to Brown's criticism that QBism faces an explanatory gap regarding the objective description of the world. Sections 4 and 5 contain the heart of the paper. In section 4, drawing upon Born's notion of invariance as the mark of the real, it is argued that there is no essential difference between the Einstein's program of 'the real' and the QBists' program. In the next section, it will be argued that QBism can account for the unitary evolution of the quantum state. Finally, in Section 6, we conclude with the findings of the paper.

## 2. Are quantum states part of the stuff of the world?

Brown claims that there are powerful plausibility arguments for the view that the quantum state is something real and,

'They almost all have to do, in one way or another, with quantum phase, with the fact that the wavefunction, in its relation to probability, is strictly a (generally complex) probability *amplitude*: it has more structure than a probability distribution does.' (Brown 2017; 5)

[2] . There is no such thing as one single "Copenhagen Interpretation", see e.g. Howard (2004).

So, according to this argument, the wavefunction has properties such as interference and entanglement which, it is claimed, to explain them one has to regard the wavefunction as something real, i.e., as describing or representing the way the world is. However, QBists maintain that quantum states are degrees of belief of an agent concerning her measurement outcomes. A quantum state is a catalogue of beliefs. But if so, there seems to be an 'explanatory deficit' in QBism (Timpson 2008). For example, consider interference phenomena. The proponents of an explanatory deficit in QBism claim that it could explain why an agent would believe in interference phenomena but could not explain why these phenomena occur. It seems that the quantum state has properties such as interference which could not be regarded as beliefs. Therefore, the quantum state is more than a subjective catalogue of beliefs. 'It has more structure than a probability distribution does'.

However, as will be shown below, the response to the objection that the quantum state is a probability amplitude and consequently more than a probability distribution lies in the fact that we could express the quantum state in terms of a pure probability distribution using minimal informationally complete (MIC) measurements (DeBrota et al., 2019; 2020). Or, we could give the Born rule in terms of pure probabilities using symmetric informationally complete (SIC) measurements (Fuchs 2017b).

Consider a quantum system with a $d$-dimensional Hilbert space $H_d$ , and suppose that there is a SIC measurement for it. Let us call its outcomes $O_i$ , $i = 1, \dots, d^2$.[3] A SIC is a set of $d^2$ rank-one projection operators $\Pi_i = |\psi_i\rangle\langle\psi_i|$ such that (Appleby et al., 2017):

$$tr(\Pi_k \Pi_l) = \frac{d\delta_{kl} + 1}{d + 1} \tag{1}$$

Such a set of operators when rescaled appropriately form a positive-operator-valued measure (POVM), i.e.,

[3] . In what follows, we use the same letter for the measurement outcome and the corresponding operator.

$$\sum_i O_i = \sum_i \frac{1}{d} \Pi_i = I \tag{2}$$

Moreover, these operators are linearly independent and form a complete basis for the space of Hermitian operators, so that the probabilities for the outcomes of a SIC measurement $p(O_i)$ determine the quantum state $\rho$. That is, if

$$p(O_i) = \frac{1}{d} tr(\rho \Pi_i) \tag{3}$$

then

$$\rho = \sum_{i=1}^{d^2} \left[ (d+1) p(O_i) - \frac{1}{d} \right] \Pi_i \tag{4}$$

Furthermore, using equation (4), we can express the Born rule directly in terms of probabilities, without ever invoking amplitudes. For a quantum state $\rho$ and a measurement $\{E_j\}$, the Born rule gives the probabilities of the measurement outcomes according to

$$Q(E_j) = tr(\rho E_j) \tag{5}$$

Substituting equation (4) in the formula (5) gives (Fuchs and Schack 2013; Fuchs 2017b):

$$Q(E_j) = \sum_{i=1}^{d^2} \left[ (d+1) p(O_i) - \frac{1}{d} \right] p(E_j | O_i) \tag{6}$$

In the above equation, we have

$$p(E_j | O_i) = tr(E_j \Pi_i) \tag{7}$$

which is a conditional probability for $E_j$ upon having obtained $O_i$. This is because an agent performing a SIC measurement on a system with state $\rho$ and then going to the measurement $\{E_j\}$, has to update from $\rho$ to $\Pi_i$ using Lüders rule.

But, one might object that there is no proof for the existence of SICs in all dimensions.[4] However, it is not fatal to the above argument as one could express the Born rule using any informationally complete POVM which are known to always exist (DeBrota and Stacey 2019). The only difference is that in this case the formula for the Born rule would be more complicated. Still, the concepts are the same. Here, it should be emphasized that for QBists it is the Born rule that is the fundamental element of quantum mechanics, not the POVMs or the quantum states. The distinctive structure of the Born rule in contrast to the classical Law of Total Probability is what gives quantum mechanics its characteristic features, notably its Hilbert space structure (Fuchs and Stacey 2019).

Moreover, why does the Born rule have such a central role in the formalism of quantum mechanics? The Born rule allows calculating probabilities using the quantum states. Conversely, assigning probabilities for any sufficiently rich set of measurements, or even for a single but informationally complete measurement as we have shown, is mathematically equivalent to assigning a quantum state itself. The two kinds of assignments determine each other uniquely (Fuchs 2010). Therefore, as Boge puts it:

‘There are reasons to suspect that QM is an *inherently probabilistic* theory, and depending on one’s interpretation of probability (or the probabilities at play, at any rate) this could again be fleshed out to mean that knowledge or information or belief. . . are at stake *in some sense* in the interpretation of QM all along.’ (Boge 2018; 290)

But, if the quantum state is epistemic and represents the beliefs of the agent assigning the quantum state, what are these beliefs about? And since the quantum state is the essential element of quantum mechanics, this amounts to the following question: what is quantum mechanics about? We turn to that in the next section.

## 3. What is quantum mechanics about according to QBism?

[4] . For recent developments regarding the existence of SICs see Bengtsson (2020).

According to a common understanding, science is about describing an external mind-independent world without any recourse to the observer. In other words, science is about an objective world by virtue of eliminating the role of the observer in the scientific enterprise. However, QBists believe that this is a misunderstanding of science 'because everything any of us knows about the world is constructed out of his or her individual private experience, it can be unwise to rely on a picture of physical world from which personal experience has been explicitly excluded, as it has been from physical science' (Fuchs et al., 2014). They argue that QBism corrects this misconception by putting the observer back into science (Mermin 2014). In particular,

'A QBist takes quantum mechanics to be a personal mode of thought – a very powerful tool that any agent can use to organize her own experience. … But quantum mechanics itself does not deal directly with the objective world; it deals with the experiences of that objective world that belong to whatever particular agent is making use of the quantum theory.' (Fuchs et al. 2014; 750)

What does it mean that quantum mechanics is a personal mode of thought? In QBism, any agent who uses quantum mechanics models all phenomena except her direct internal awareness of her private experience according to it. To do this, she relies on her past experience to assign quantum states to her systems of interest. The quantum states are probability assignments which express the agent's probabilistic expectations for her future experience. From the QBist's subjective view on probabilities it follows that the agent's probability assignments express her degrees of belief about her future experience.

Although, Brown shares the subjective interpretation of probability with QBism, he does not accept

'the further inference in QBism that our scientific reasoning should primarily be about personal experiences, our "beliefs", and not the objective world'. (Brown 2017; 15)

Brown's examples of what constitutes a description of the objective world are interesting and deserve attention in their own right for what follows. He mentions the explanation of what has actually happened in the early universe and what has actually been happening inside the stars since they are examples of what science should try to tell us about the objective world. According to QBism, what science has to tell us on these matters are models constructed to account for current astrophysical data and observations. But, Brown claims that here 'we seem to be left with an explanatory gap' (ibid.). He means that the QBist's answer leaves unexplained what was going on in the early universe or has been going on inside the stars. From this viewpoint, science should give us a description of the world according to what Pauli calls 'the ideal of the detached observer':

'To put it drastically the observer has according to this ideal to disappear entirely in a discrete manner as hidden spectator, never as actor, nature being left alone in a predetermined course of events, independent of the way in which the phenomena are observed.' (Pauli, as quoted in Fuchs 2017b; 23)

Interestingly, Pauli notes that the historical origin of this ideal goes back to celestial mechanics. Note that Brown's examples are also taken at least partly from celestial mechanics.

However, as Pauli emphasizes, quantum mechanics forces us to abandon this ideal of the detached observer. This is formally expressed by the reduction of wave packet upon measurement. Accordingly, there is an inevitable and unpredictable change of the state of the system upon measurement. This for Pauli implies the abandonment of the idea of the detachment of the observer from the course of physical events outside herself. Why? It is because for Pauli as for QBists the measurement instruments are literally extensions of the agent. In order to illustrate the point of the argument, let us first look at what Bohr has to say about the ideal of the detached observer:

'We have in quantum physics attained the same goal [of keeping the ideal of the detached observer] by recognizing that we are always speaking of well defined observations obtained under specified experimental conditions. These conditions can be communicated to everyone who also can convince himself of the factual character of the observations by looking on the permanent marks on the photographic plates.' (Bohr, as quoted in Fuchs 2017b; 23)

So, according to Bohr's argument, the detachment of the observer from the course of physical events outside herself is secured through the elimination of the subjective elements in the account of experience. That is, the experimental arrangement is independent from the whim of the agent in the sense that its factual character could be checked by everyone by looking at the permanent marks obtained through the experiment.

But, as Pauli emphasizes, reference to experimental conditions is 'information on the observer' and establishment of an experimental arrangement is an 'action of the observer' (Fuchs 2015; 46). As Fine (1986; 155) puts it, 'the probabilities of the theory are generally understood as probabilities for various measurement outcomes and, so understood, suppose a prearranged apparatus for measurement – and hence a measurer-observer of some sort'. Reference to factual character of the observations obtained under well defined experimental conditions could eliminate the role of this or that observer but in no way could eliminate the subjective element in general. The establishment of the experimental arrangement is in a sense the embodiment of the observer.

According to QBism, the experimental apparatus are the extensions of the agent in the sense that the outcomes of the experiments carried out by those apparatus are personal experiences of that agent. That is why Bohr (1949) insists on using 'unambiguous language with suitable application of the terminology of classical physics' to express the set-up and outcomes of the experiments. As Mermin puts it:

'Ordinary language was enormously important to Bohr. An essential part of an experiment was reporting it to others. I always found this puzzling. If the outcome of an experiment is an objective classical fact, why is it so important to be able to communicate it to other people in ordinary language? Why can't they just look for themselves?' (Mermin 2018; 20)

The QBists' answer is that since the outcomes of the experiments are the personal experiences of the agent using an unambiguous language is the only way to communicate them. For QBists, "Wigner's friend" (Wigner 1961) shows that we should recognize as fundamental rather than paradoxical the outcomes of experiments being personal experiences of the agent carrying out the experiment. In Wigner's friend scenario, Wigner and his friend assign different quantum states to the same quantum system. Now the question is this: which one is correct? If the outcome of an experiment is an objective classical fact available for anyone's inspection why should Wigner and his friend assign different states to the same system? One might answer that they have different states of information. But, who has the right state of information? According to QBism, the whole puzzle stems from insisting on taking what are personal experiences as agent-independent objective facts.

Moreover, as Ihde (1991) argues, the experimental instruments are not neutral windows on the world-in-itself. They are 'hermeneutic devices' that prepare and make readable what we call scientific objects (Ihde 1998; 149). Through the mediation of experimental instruments, our experience is already of an interpreted reality.

Now we are ready to answer Brown's criticism that QBism faces an explanatory gap regarding the objective description of the world. He claims that the question '*what* was evolving in the early universe, if not quantum states?' leads nowhere in QBism. Because according to QBism, quantum states are probability assignments of agents and there were no agents in the early universe to assign them. However, in the light of the above discussion, Brown's criticism suffers from a misconception about science. It is exactly the misconception that QBism seeks to correct. Science is not a neutral description of a mind-independent world. All of our descriptions of the past and future of the world are nothing but our beliefs based on current experimental data and observations. And all these data and observations are obtained through experimental conditions that are without exception instrumentally-mediated which in turn means that we are already experiencing an interpreted reality. The experimental arrangement is the extended embodiment of the agent. Referring to the objective character of the experimental arrangements and the results obtained based on them to secure the elimination of the subjective element is doomed to failure. This maneuver could at best eliminate the role of particular agents but not the agent in general. One could say that our scientific facts are the beliefs of the scientific community. That is, they are types of beliefs that could withstand intersubjective testing. To the question 'what was going on in the early universe?' the only answer is that what 'we say about the past' is really a statement about what we expect for our futures. This belief is not and could not be about the early universe itself rather it concerns our model to account for current data and observations. Beyond and behind this conceptual model we have access to nothing in principle.[5]

The charge of the existence of an explanatory gap in QBism stems from a particular conception of the nature of physical theories that QBists are not committed to. According to this representationalist view, theories must or do *represent* reality as it is in itself. Consequently, there is an explanatory gap in QBism because the quantum states represent degrees of belief of the agents and not reality. But, quantum theory is about the outcomes of the actions or interventions of the agents on the world. As Finkelstein (1996) puts it, 'quantum physics is action physics'. A quantum state is a state of belief about the outcomes of the agent's actions on the quantum system rather than a third-person representation of a reality in itself. This issue of representation and reality deserves more attention and will be considered more fully in the following section.

## 4. Einstein's Realism and QBists' Realism

Regarding the relation between QBism and Einstein's philosophy of science, Fuchs points out that he 'cannot see any way in which the program of QBism has ever contradicted what Einstein calls the program of "the real".' (Fuchs 2017a; 118)

However, Brown argues that there is an essential difference between the program of QBism and Einstein's program. According to QBists,

'… the best understanding of quantum theory is obtained by recognizing that quantum states, quantum time-evolution maps, and the outcomes of quantum measurements all live within what Einstein calls the subjective factor.' (ibid., 119)

[5] . For more on this point see Section 4 and also Peierls (1991).

Brown claims that this is antithetical to Einstein's program because it leaves the objective factor out. Brown believes that for Einstein, quantum states are probability distributions over hidden variables and these are the objective factor, i.e. correspond to concepts independent of perception.

However, it is not the case that there is no objective factor in the QBists' program. For QBists the objective factor is the quantum system which is a "real existence" external to the agent (Fuchs 2010). This point notwithstanding, Brown argues that the nature of the external world of QBism is 'ineffable' and so something of little interest to Einstein:

'The external physical systems float free of the quantum formalism. No *describable* objective attributes can be assigned to these systems in QBism, because, as we have seen, the universe is made of something other than quantum states, and quantum states are all we have in the formalism of quantum mechanics.' (Brown 2017; 21)

But, we think this argument is implausible. According to QBism, quantum mechanics deals with the experiences of the agents using quantum mechanics. In this view, the notion of an agent-independent reality is constructed through intersubjective testing of the experiences of agents. Different agents could communicate their experiences to each other and through intersubjective testing 'a common body of reality can be constructed' (Fuchs et al., 2014). What we call 'the world' is constructed from our intersubjective experiences. The quantum state represents 'one's probabilistic expectations for one's subsequent experience'. This is the scope of our knowledge. And we use this knowledge to find our way through the maze of experience. As will be shown below, we do not think Einstein's program is any different.

In order to assess the plausibility of Brown's criticism it is necessary to see what Einstein's program of "the real" is. For Einstein,

'The aim of science is, on the one hand, a comprehension, as *complete* as possible, of the connection between the sense experiences in their totality, and, on the other hand, the accomplishment of this aim *by the use of a minimum of primary concepts and relations*.' (Einstein 1936; 352)

In other words, according to Einstein's program of the real, in physics we take the existence of sense impressions or experiences as given and try to understand the connections between them by constructing a conceptual system. This conceptual system has a layered structure. At the lowest level lie primary concepts which are 'directly and intuitively connected with typical complexes of sense experiences'. However, this first layer lacks logical unity. In order to supply our conceptual system with logical unity we invent concepts and relations which are only indirectly connected with sense experiences, i.e. retaining the primary concepts and relations as logically derived concepts and relations. Striving for more unity, we invent a tertiary level of concepts and relations for deduction of the concepts and relations of secondary and so indirectly of the primary layer. This process goes on until a conceptual system with greatest unity is achieved which is compatible with our sense experiences. We assume this aim in physics but there is no guarantee that it would result in a definite system.

So, what is the real or the objective factor in Einstein's program? It has to do with *sense impressions* because everything else is the free creation of human mind. Accordingly, Einstein (1949) defines the objective factor as the totality of those concepts and relations which are independent of the act of perception. But, what does it mean for a totality of concepts and relations to be independent of the act of perception? We think for an answer we should turn to what Max Born has to say about the concept of reality in physics:

'It presupposes that our sense impressions are not a permanent hallucination, but the indications of, or signals from, an external world which exists independently of us. Although these signals change and move in a most bewildering way, we are aware of objects with invariant properties. The set of these invariants of our sense impressions is the physical reality which our mind constructs in a perfectly unconscious way. This chair here looks different with each movement of my head, each twinkle of my eye, yet I perceive it as the same chair. Science is nothing else than the endeavour to construct these invariants where they are not obvious.' (Born 1949; 103-4)

Einstein advocates a similar view:

'By the aid of speech different individuals can, to a certain extent, compare their experiences. In this way it is shown that certain sense perceptions of different individuals correspond to each other, while for other sense perceptions no such correspondence can be established. We are accustomed to regard as real those sense perceptions which are common to different individuals, and which therefore are, in a measure, impersonal. The natural sciences, and in particular, the most fundamental of them, physics, deal with such sense perceptions.' (Einstein 1922; 1)

Therefore, as Born (1953) emphasizes, 'the idea of invariant is the clue to a rational concept of reality'. Similarly, according to QBism, the invariant in quantum mechanics is the Born rule. All agents should strive to form their probabilistic expectations of the consequences of their measurements such that the Born rule is satisfied. Violating the Born rule would result in disastrous consequences. Quantum mechanics grasps the real through the Born rule.

So, we think what Einstein calls the 'objective factor' corresponds to what Born calls the 'invariants of our sense impressions'. Both are defined by independence from the act of perception. For Einstein, the objective factor is what justifies the construction of conceptual systems, i.e. differentiates them from empty logical schemes. But, why for Einstein is not the objective factor the sense impressions themselves? It is because the sense impressions by themselves lack invariance. By constructing concepts and relations between them we achieve a degree of invariance and thereby

'We are able to orient ourselves in the labyrinth of sense impressions. These notions and relations, although free statements of our thoughts, *appear to us as stronger and more unalterable than the individual sense experience itself*, the character of which as anything other than the result of an illusion or hallucination is never completely guaranteed.' (Einstein 1936; 350, emphasis added)

The important point, as Fine (1986) emphasizes, is that in Einstein's program of the real there is nothing (such as an external reality) that stands 'outside' the conceptual system to which our concepts could be compared.[6] We are situated in 'the labyrinth of sense impressions' and it is the conceptual system that tells us what the reality is. As Einstein in a letter to Schrödinger in 19 June 1935 points out:

'The real difficulty lies therein that physics is a kind of metaphysics; physics describes "reality". But we do not know what "reality" is; we only know it through physical description!' (von Meyenn 2011, 537)

To sum up, in both the QBists' program and Einstein's program, the aim of science is to guide us through the maze of experience and accordingly the sole justification for our theories or conceptual systems is their *success* in doing so (Fuchs 2017a; Einstein 1936). We do not see any essential difference between the two programs.

However, one might object that according to Einstein's program of the real we do know reality through physical description whereas on QBism that remains to be seen. But, here also there is no essential difference between the two programs. According to QBism, physical theories are not 'attempts to directly represent (map, picture, copy, correspond to, correlate with) the *universe*—with "universe" here thought of in totality as a pre-existing, static system; an unchanging, monistic something that just *is*' (Fuchs 2010). As Howard and Giovanelli (2019) emphasize, for Einstein, realism 'is not a philosophical doctrine about the interpretation of scientific theories or the semantics of theoretical terms. For Einstein, realism is a physical postulate.' This physical postulate is spatial separability, i.e., physical systems situated in different parts of space should (in theory) have an existence independent of each other. Finally, Einstein evaluates the prospects for the success of the program of the real thus:

[6] . See also Howard and Giovanelli (2019) especially Section 5.

‘Will this credo survive forever? It seems to me a smile is the best answer.’ (Einstein 1950; 758)

That is, whether we could know reality through physical description in the ideal limit in which our theorizing finally result, remains to be seen.

## 5. The Problem of the Evolution of the Quantum State

Hitherto, we have discussed the epistemic nature of the quantum state and its consequences in the static case i.e. at a given time. In this section, we will discuss the time evolution of the quantum state according to QBism.

According to QBism, quantum states are probability assignments of the agent concerning her expectations for the outcomes of measurements on the quantum system. Brown (2017) argues that QBism has no resources to account for the evolution of the quantum state. Suppose the quantum system evolves freely over an interval of time without any measurements taking place. It follows, Brown claims, that the probability assignments of the agent does not change during this interval because there are no measurements and consequently the agent does not receive new information to update her probability assignments. That is, in QBism there is no evolution of the quantum state during the interval between measurements. However, we know that the quantum state evolves during this interval according to the Schrödinger evolution. Brown criticizes this with the following comment,

‘It is as if von Neumann’s two motions in quantum mechanics have reappeared in a different guise! The difference now is that the mystery lies with the unitary evolution.’ (Brown 2017; 17)

But, we do not think that Brown’s argument is justified. In QBism, quantum states and evolutions along with the outcomes of measurements live on the subjective factor. According to QBism, just as quantum states are personal judgments the unitary time evolutions are personal judgments too. To see this, let us again consider equation (6) of Section 2. This equation expresses the agent’s probabilities under the assumption that the SIC measurement is not performed. In other words, it is a conditional probability expressing a counterfactual situation. That is, the probabilities for the actual measurement outcomes $\{E_j\}$ depend on the probabilities for the hypothetical SIC measurement outcomes $p(O_i)$. Now, let us make the actual measurement a unitarily rotated version of the SIC measurement. Then,

$$E_j = \frac{1}{d} U \Pi_j U^\dagger \tag{8}$$

which implies a simplification of equation (6) to,

$$Q(E_j) = (d+1) \sum_{i=1}^{d^2} p(O_i) p(E_j | O_i) - \frac{1}{d} \tag{9}$$

for the probabilities of the actual measurement outcomes (Fuchs and Stacey 2019). Equation (9) shows that the unitary time evolution $U$ is a variant of the Law of Total Probability.[7] This is because in the Schrödinger picture one could regard $p(O_i)$ and $Q(E_j)$ as the SIC representations for the initial and final quantum states under the evolution $U$. It should be emphasized that all the terms in equation (9) refer to the probability assignments that the agent makes at a single time, although these assignments are about the outcomes of a hypothetical SIC measurement that could happen at two different times in the future. So, the unitary time evolution $U$ is the personal judgment $p(E_j | O_i)$. Furthermore, it resolves the issue of why there should be two kinds of state evolution, one upon measurement and the other due to unitary evolution. As Fuchs (2010; 13) puts it, 'there are not two things that a quantum state can do, only one: Strive to be consistent with all the agent's other probabilistic judgments on the consequences of his actions, factual and counterfactual'.

Thus, unitarity is a consistency constraint guiding the agent on how to relate her beliefs about a measurement to be performed at different circumstances namely, a measurement that could be performed either at time $t_1$ or at time $t_2$ in the

[7] . See also Stacey (2016).

agent's future. These beliefs are held by the agent simultaneously at a definite time. So, the unitary time evolution formula

$$\rho^{'} = U(t)\rho U^{\dagger}(t) \qquad (10)$$

guides the agent on how to relate her probability assignments for a measurement at $t_1$ to her probability assignment for the same measurement at $t_2$ using the unitary map $U(t)$.

More importantly, the real mystery with regard to the evolution of the quantum state is this: why unitarity in the first place? It has all to do with probabilities. As Hughes (1989, 115) illustrates if we 'consider a state just as a probability function on a set of experimental questions' along with a few other plausible assumptions we could derive the unitary structure of quantum theory. Further in this direction, Appleby et al. (2017) show that the unitary symmetry specifies all of the defining features of quantum theory. But why this symmetry group in particular? There is no satisfactory answer yet. That is, we know that regarding quantum states as 'catalogues of expectations' and assuming unitarity one could reconstruct quantum theory. Formally, these two assumptions are our *axioms*.

## 6. Conclusion

In this paper, we have argued for the following claims. First, it is argued that the charge of the existence of an explanatory gap in QBism regarding the objective description of the world stems from a particular conception of the nature of physical theories that QBists are not committed to and seek to correct. According to this representationalist view, theories must or do *represent* reality as it is in itself. However, for QBists, quantum theory concerns what we expect from our actions or interventions on the world rather than a third-person representation of it. Moreover, this provides an answer to Brown's 'problem of the past'. That is, how can QBism account for events in the past history of the world? According to the QBist conception of scientific theories, what we say about the past ultimately concerns what we expect about the future. Our scientific beliefs about the past concern our conceptual models to account for the current data and observations.

Second, it is argued that there is no essential difference between Einstein's program of the real and the QBists' program. Drawing upon Born's notion of invariance as the mark of the real, it is argued that both programs do regard and try to discover invariances as what is real or objective. Moreover, in both programs, the aim of scientific theories is to guide us through the maze of experience and accordingly the sole justification for our theories or conceptual systems is their *success* in doing so. Third, it has been shown how in QBism, just as quantum states are personal judgments, so it is with unitary time evolutions. Contra Brown, there is nothing more mysterious about unitary evolution than quantum collapse (i.e., state update due to measurement) or even the Born rule itself: they all have the character of normative principles. QBism is not so simply dismissed.

**Acknowledgements**: I especially thank Chris Fuchs for many valuable comments and suggestions. I'm also grateful to Florian Boge, Jacques Pienaar, Blake Stacey and John DeBrota for helpful comments on earlier drafts of this paper. I'm also grateful to Ebrahim Azadegan.